\documentclass{aa}  

\usepackage[varg]{txfonts}
\usepackage{graphicx}
\usepackage{txfonts}
\usepackage{lipsum}
\usepackage{subcaption}         
                                
\usepackage{lscape}             
                                
\usepackage{placeins}           

\usepackage{amsmath}
\usepackage{aas_macros}

\usepackage[colorlinks, citecolor=black, linkcolor=blue, urlcolor=blue]{hyperref}
\usepackage{xcolor}

\newcommand{\RomanNumeralCaps}[1]
    {\MakeUppercase{\romannumeral #1}}
    
\begin{document}

\title{Cosmological implications of the Gaia Milky Way declining rotation curve.
}

\author{E. Coquery\inst{1}
\and A. Blanchard\inst{2}}

\institute{{\'E}cole Centrale de Lyon, 36 Avenue Guy de Collongue, 69134 Écully, France\\
\email{even.coquery@gmail.com}
\and Universit{\'e} de Toulouse, UPS-OMP, IRAP, CNRS, 14 Avenue Edouard Belin, F-31400 Toulouse, France\\
\email{alain.blanchard@irap.omp.eu}}

\abstract
{Although the existence of dark matter has been widely acknowledged in the cosmology community, it is as yet unknown in nature, despite decades of research, which questions its very existence. This never-ending search for dark matter leads to consider  alternatives. Since increasing the enclosed mass is the only way to explain the flat appearance of galaxies' rotation curves in a Newtonian framework, the MOND theory proposed to modify Newton's dynamics when the acceleration is around or below a threshold value, $a_0$. 
Observed rotation curves, generally flat at large distances, are then usually well reproduced by MOND with $a_0 \sim 1.2\times10^{-10}$ m/s$^{2}$. However, the recent Gaia evidence of a decline in the Milky Way rotation curve is a distinct behavior.
Therefore, we examine whether MOND  can accommodate the Gaia declining rotation curve of the Milky Way. 
 We first depict a standard model to describe the Milky Way's baryonic components. Secondly, we show that a NFW (Navarro, Frenk, \& White ) model is able to fit the decline, assuming a scale radius $R_s$ of the order of $4$ kpc. In a third step, we show that the usual MOND paradigm is not able to reproduce the declining part for a standard baryonic model. Finally, we examine whether the MOND theory can accommodate the declining part of the rotation curve when relaxing the characteristics of the baryonic components. To do so 
  we use a MCMC method on the characteristics of the stellar and the $\text{H\RomanNumeralCaps{1}}$ disk, including their mass. 
  We found that the stellar disk should be massive, of the order of $10^{11}$ M$_{\odot}$. The $\text{H\RomanNumeralCaps{1}}$ disk mass is capped at nearly 1.8 $\times 10^{11}$ M$_{\odot}$ but could also be negligible. Finally, $a_0$ is consistent with 0, with an upper limit of $0.53\times10^{-10}$ m/s$^{2}$ (95\%), a value much lower than the above mentioned value usually advocated to explain standard flat rotation curves in MOND theory.}

\keywords{Suggested keywords}
\maketitle

\section{Introduction}

\noindent Studying galaxies' rotation curves (RC) is of paramount importance in cosmology, as they hint towards the existence of dark matter. Indeed, observations have shown that most galaxies' rotation curves are flat, which is in strong disagreement with Newtonian dynamics' predictions for known baryonic components. Historically, cosmologists' favorite answer to this crisis is to assume the existence of an invisible mass in order to accommodate the observations. This assumption is also supported by the observations of unrelated phenomena, such as 
gravitational lensing \citep{Refregier_2003}, cosmological microwave background \citep{2016}, and more. Dark matter, whose quantity was estimated to be nearly 30\% of the Universe mass-energy content, has now become a pillar of cosmology's standard model, $\Lambda$CDM.

However, since dark matter has never been observed directly, we could alternatively envisage that Newtonian dynamics may change under certain conditions. The MOND\footnote{Which stands for MOdified Newtonian Dynamics} theory \citep{1983ApJ...270..365M} 

MOND proposes a modification of Newtonian dynamics in regimes where the acceleration approaches or drops below a characteristic value of the acceleration
$a_0$. In the deep MOND limit, the rotation velocity for a circular orbit resumes to $v=(GMa_0)^{1/4}$, which explains the flat appearance of most rotation curves. Previous fits of rotation curves using MOND agree on a value of about $1.2\times10^{-10}$ m/s$^2$ \citep{begeman_extended_1991}, an acceleration threshold so low that MONDian effects could not be detected on Earth nor in the Solar System. This theory has been further developed, leading to more complex formulations such as AQUAL \citep{1984ApJ...286....7B}, QUMOND \citep{milgrom_generalizations_2023}, or 
the relativistic TeVeS \citep{2004PhRvD..70h3509B}. However, these theories face tensions in multiple fields, such as galaxy cluster dynamics \citep{Sanders_2003}, CMB anisotropies, or matter power spectrum \citep{DODELSON_2011}. To alleviate these tensions, hybrid theories have been built (see for example \cite{Bruneton_2009} or \cite{Skordis_2021} for $\Lambda$CDM cosmology with MONDian effects at galactic scales). Despite those tensions, MOND remains a 
simple theory for reproducing flat galaxies' rotation curves{, although there are claims that the agreement does not favor MOND from a statistical analysis of Spitzer Photometry and Accurate Rotation Curves (SPARC) galaxies' rotation curves \citep{2024arXiv240110202K}}.

Although it is relatively  easy to measure other galaxies' rotation curves \citep{Corbelli_2000,1975IAUS...69..331R}, probing the Milky Way's from the inside is a challenge at high radius. Previous data tend to confirm that the MW's RC is flat \citep{mroz_rotation_2019}, consistent  with both dark matter and MOND , although some moderate  decline has been reported \citep{robin_self-consistent_2022} in agreement with earlier claim  consistent with Radial Acceleration Relation (RAR) \citep{2019ApJ...885...87M}. A moderate decline was also inferred for the most luminous galaxies \citep{1996MNRAS.281...27P,2007MNRAS.378...41S}. However, Gaia's latest data release \citep{vallenari_gaia_2022} sheds light on a velocity decline after $\approx 15$ kpc,  of the order of 3.5 km/s/kpc,  which different analyses agree on (see for example \cite{wang_mapping_2022}, \cite{zhou_circular_2023}, \cite{jiao_detection_2023},  {\cite{2024MNRAS.528..693O}}).   This decline is also consistent with previous indications for the presence of a dip in the Milky Way's RC \citep{2016MNRAS.463.2623H} with a higher significance level.
 Some other galaxies' rotation curves were also found to be declining : a sample of twenty-two galaxies was
studied under the MOND paradigm by \cite{Zobnina_2020}. They conclude that some galaxies' rotation curves do not meet with the usual MOND paradigm, needing a  value of $a_0$ 
 lower than values obtained from previous fits of the rotation curves. Studying Gaia's newfound decline might yield similar conclusions in the Milky Way. Milky Way rotation curve measurements before the decline are nonetheless more scattered, as several papers account for very different values (e.g. \cite{2015NatPh..11..245I,mroz_rotation_2019,zhou_circular_2023,labini_mass_2023,wang_mapping_2022,mcgaugh_precise_2018} and  \cite{robin_self-consistent_2022}).  
Such  considerations lead us to focus on the declining part of the RC, without considering the inner part.

In this paper, we aim at shedding light on the declining RC's implications regarding MOND   as inferred from Gaia, without attempting to fit the inner part. We start by building a model for the baryonic components of the Milky Way, in order to compute the rotation curve at various radii from the galactic center. In order to compare $\Lambda$CDM with MOND, we implement this model both under a dark matter paradigm and under Milgrom's modified dynamics. We then use this model as a basis to find the optimal value of $a_0$ that fits the decline - if such a value exists. Finally, we compare MOND results with $\Lambda$CDM, and detail why the MOND paradigm 
does not accommodate the Milky Way rotation curve under reasonable assumptions.   This interpretation of the data is based on the assumption of simple dynamics that are not violently disturbed in the outer parts of the Galaxy \citep{2024A&A...692A..50K,2024ApJ...970...94K}.\\

\section{The Milky Way's rotation curve}\label{sec2}

\subsection{Modeling the baryonic components of the Milky Way} \label{s:bcmw}

\noindent In order to compute the rotation curve, we first need to establish a model for the spatial distributions of the baryonic components of the Milky Way. 
We choose to work with the B2 model used in \cite{jiao_detection_2023}, and described in \cite{de_salas_estimation_2019}. This model consists of three components :
\begin{itemize}
    \item A spherical bulge modeled by a Hernquist potential : 
    \begin{equation}
    \Phi(r)=-\frac{GM_{bulge}}{r+r_b}
    \label{eq:Hpot}
\end{equation}

    \item A thin stellar disk,
    \item Multiple gas disks.
\end{itemize}

Judging by the values provided by \cite{de_salas_estimation_2019} and references therein, we only consider the \text{H\RomanNumeralCaps{1}} disk, since the other gas masses are negligible.\\

Both the stellar disk and the \text{H\RomanNumeralCaps{1}}  disk densities are given by a double-exponential :
\begin{equation}
        \rho^i (R,z) = \rho_0^i \exp(-\frac{R}{r_d^i}-\frac{|z|}{z_d^i})
    \label{eq:B2}
\end{equation}
where $i\in \{st.,$ \text{H\RomanNumeralCaps{1}}\}.

$\rho_0^i$ is a normalization constant :

\begin{equation*}
  \left\{
    \begin{aligned}
      & \rho_0^{st.}=\frac{M^{st.}}{4\pi (r_d^{st.})^2 z_d^{st.}} \\
      & \rho_0^{\text{H\RomanNumeralCaps{1}}}=\frac{M_{\text{H\RomanNumeralCaps{1}}}}{4\pi r_d^{\text{H\RomanNumeralCaps{1}}} z_d^{\text{H\RomanNumeralCaps{1}}}(R_t+r_d^{\text{H\RomanNumeralCaps{1}}})e^{-R_t/r_d^{\text{H\RomanNumeralCaps{1}}}}}\\
    \end{aligned}
  \right.
\end{equation*}
where $M^{i}$, $M_{bulge}$, $r_d^i$, $z_d^i$, $R_t$ and $r_b$ can be found in \cite{de_salas_estimation_2019}, and summed up in Table \ref{b2_params}.

\begin{table}[ht]
    \centering
    \caption{B2 model parameters \citep{de_salas_estimation_2019}}
\begin{tabular}{cc}
\hline
Parameter                 & Value (B2 Model)    \\ \hline \vspace{0.13cm}
$M^{st.}$ (M$_{\odot}$)   & $3.65\times10^{10}$ \\ \vspace{0.13cm}
$r_d^{st.}$ (kpc)         & 2.35                \\ \vspace{0.13cm}
$z_d^{st.}$ (kpc)         & 0.14                \\ \vspace{0.13cm}
$M^{\text{H\RomanNumeralCaps{1}}}$ (M$_{\odot}$)    & $8.2\times10^9$     \\ \vspace{0.13cm}
$r_d^{\text{H\RomanNumeralCaps{1}}}$ (kpc)          & 18.24               \\ \vspace{0.13cm}
$z_d^{\text{H\RomanNumeralCaps{1}}}$ (kpc)          & 0.52                \\ \vspace{0.13cm}
$R_t$ (kpc)               & 2.75                \\ \vspace{0.13cm}
$M_{bulge}$ (M$_{\odot}$) & $1.55\times10^{10}$ \\ \vspace{0.13cm}
$r_b$ (kpc)               & 0.7                 \\ \hline
\end{tabular}
\label{b2_params}
\end{table}

The stellar population of the Galaxy can be described by more complex models with several different components. The presence of a thick disk for instance impacts  the vertical structure of the velocity dispersion as well as the possible flaring of the disk \citep{2025ApJ...978...45L}. In order to check for the implication of the presence of a thick disk we also compute the circular velocity in a model  comprising two equal mass components \citep{2017A&A...598A..66P}, one in a thin disk and one in a thick disk following the Miyamoto–Nagai profile \citep{1975PASJ...27..533M}. From figure 1 and 2, one can see that the rotation curve is nearly identical. Concerning the flaring, we note that \cite{2024ApJ...976..185S} concluded that its impact has only a marginal effect on the RC at large distances, i.e. at scales relevant to our study. These conclusions are  likely due to the fact that in our work the RC is examined at a distance far away from the bulk of the stellar components. 

\subsection{Rotation curve under $\Lambda$CDM}

\noindent In Newtonian dynamics, assuming a circular movement, each mass component produces an acceleration field that can be written in terms of a circular velocity associated to this component :
\begin{equation}
    \frac{(v^i)^2}{r}=a=K_r^{i}
    \label{v_circular}
\end{equation}
where  {$a$ is the radial acceleration,} $K_r^{i}$  {is the total radial force per unit mass}, and $i$ corresponds to any baryonic component described above (stellar disk, bulge, gas disk) or to the added dark matter mass component : $i \in \{st., bulge,$ \text{H\RomanNumeralCaps{1}}$, dm \}$.

Which means that in order to compute the velocity, we first have to determine the radial force. We do so by integrating Poisson's equation with the density $\rho^i$. For the density given in Equation \ref{eq:B2}, Poisson's equation can be solved in terms of Hankel transforms. We can thus find the radial force by considering the R-derivative. Details of calculation can be found in \cite{kuijken_mass_1989}. 
We can easily determine $v_{bulge}$ using Equation \ref{v_circular} and the first derivative of Hernquist's potential  \ref{eq:Hpot}.

Finally, we need to specify a dark matter model to compute $v_{dm}$. In $\Lambda$CDM dark matter potentials are expected to follow a NFW profile \citep{1996ApJ...462..563N}. We will therefore use a standard NFW model \citep{navarro_structure_1996}

\begin{equation}
    \rho_{NFW}(r) = \frac{\rho_{0,NFW}}{r/R_s(1+r/R_s)^2},
    \label{nfw_density}
\end{equation}
where $\rho_{0,NFW}$ and the scale radius $R_s$ are free parameters. Then we only need to integrate the density of Equation \ref{nfw_density} between $0$ and a given radius $r$ to get the enclosed mass in the corresponding sphere, and then use the circular movement assumption to obtain the velocity~:

\begin{equation*}
    v_{dm}(r)^2 = \frac{{GM_{dm}(r)}}{r}.
\end{equation*}
The rotation curve under circular movement assumption can now be evaluated :

\begin{equation}
    v^2 = v_{st.}^2+v_{gas}^2+v_{bulge}^2+v_{dm}^2.
    \label{total_velocity}
\end{equation}

\begin{figure}[ht]
    \centering
    \includegraphics[width=1\linewidth]{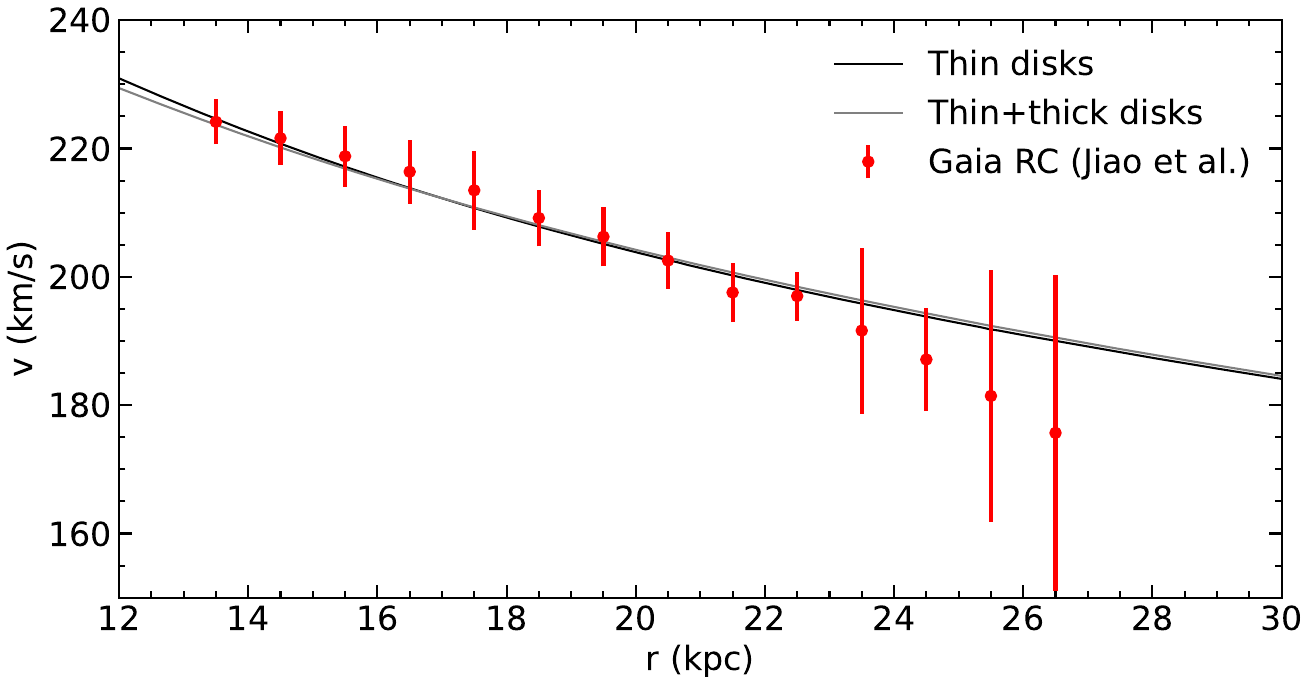}
    \caption{Rotation curve decline example using NFW {{}with} $\rho_{0,NFW}=1.8\times10^8$ M$_{\odot}$/kpc$^3$, {{$R_s^{thin}=3.97$ kpc, $R_s^{thin+thick}=3.98$ kpc.}}  The parameters for the baryonic components are  those of the B2 model in Table \ref{b2_params} taken from 
    \cite{jiao_detection_2023}. We also provide the rotation curve (grey lines) with two disks, one thin and one thick (see section 2.1).}
    \label{fig:nfw_decline}
\end{figure}

Using Equation \ref{total_velocity}, we can now compute the rotation curve under the  $\Lambda$CDM dark matter paradigm. We provide an example in Figure \ref{fig:nfw_decline}, with parameters chosen to yield an acceptable fit to the declining part of the rotation curve. This example shows that NFW is able to accommodate the declining part, with a $\chi^2$ of {{$\chi^2_{thin}=2.6$, $\chi^2_{thin+thick}=3.1$}} {({reduced $\chi$  of 0.2 and 0.24)} }which is  {{low, but }} acceptable.

This yields different results than \cite{jiao_detection_2023}.
When computing the dynamical mass using the same critical density value as \cite{jiao_detection_2023} using the parameters from our fit of the RC under NFW, we get $M_{dyn}=4.28\times10^{11}$ M$_{\odot}$ for a virial radius of 153.5 kpc. These values are higher than the results from \cite{jiao_detection_2023}  {and consistent with \cite{2024ApJ...976..185S}}. However, they remain consistent with their newfound upper limit on the dynamical mass. In addition, extrapolation to a distance nearly ten times greater is certainly very uncertain : on galaxy scale strong feedback are expected \citep{1992A&A...264..365B}, which could seriously alter the dark matter distribution in the inner part of galaxies \citep{2022A&A...665A.143L}. The concentration parameter $c$ for the above characteristics is  larger than found by \cite{2019ApJ...871..120E}, but our fit is performed only on $R > 13$ kpc. Our value of $c$ is much larger  than \{bf{average value} expected from LCDM simulations involving only dark matter \citep{2001MNRAS.321..559B}. The role of baryons may however considerably enhance the concentration parameter on galaxy scale \citep{2023MNRAS.523.3258S}. We notice  that \cite{2019MNRAS.487.5679L} infer a determination of the Milky Way's  rotation curve up to 100 kpc, consistent with a NFW profile,  with data roughly consistent with the ones  used in Figure \ref{fig:nfw_decline}, although not reproduced by the global fit which covers the  RC in the range $4.6-98$ kpc  from \cite{2016MNRAS.463.2623H}. Finally, we note that dynamical arguments from the stellar streams dynamics \citep{2024ApJ...967...89I} as well as from the local group kinematics \citep{2024A&A...689L...1B} pointed consistently towards Milky Way mass of the order or above $10^{12}$ M$_{\odot}$ .

\subsection{Rotation velocity under MOND}

Using the standard MOND formulation instead of a more rigorous one on a rotation curve computation, leads to a difference of about 5\% \citep{lopez-corredoira_exact_2021}. For the sake of simplicity, we will thus stick to Newton's second law as modified in \cite{milgrom_mond_2015} :

\begin{equation}
    \mu(a/a_0)a=K_r
    \label{mond_equation}
\end{equation}
where  {$a$ is the acceleration that now may differ from } $K_r$  the total radial force per unit mass and $\mu$ is a function chosen such as \citep{mcgaugh_mass_2004} :
\begin{equation*}
  \left\{
    \begin{aligned}
      & \lim_{a>>a_0} \mu(a/a_0) = 1 \\
      & \lim_{a<<a_0} \mu(a/a_0) = a/a_0.\\
    \end{aligned}
  \right.
\end{equation*}

In the following, we choose to work with $\mu(x)=\frac{x}{\sqrt{1+x^2}}$, as it is standard and used by \cite{mcgaugh_mass_2004}\footnote{This choice is not important as we focus on the asymptotic behavior of the curve.  {We have checked that the flattening of the rotation curve is similar with the so-called simple function \citep{2005MNRAS.363..603F}. }}. By only keeping the $a>0$ solution, one can derive the total velocity from Equation \ref{mond_equation} :
\begin{equation*}
    v(r)=\left(\frac{r^2}{2}\left(K_r^2+\sqrt{K_r^2(K_r^2+4a_0^2)}\right)\right)^{1/4}.
\end{equation*}

\begin{figure}[ht]
    \centering
    \includegraphics[width=\linewidth]{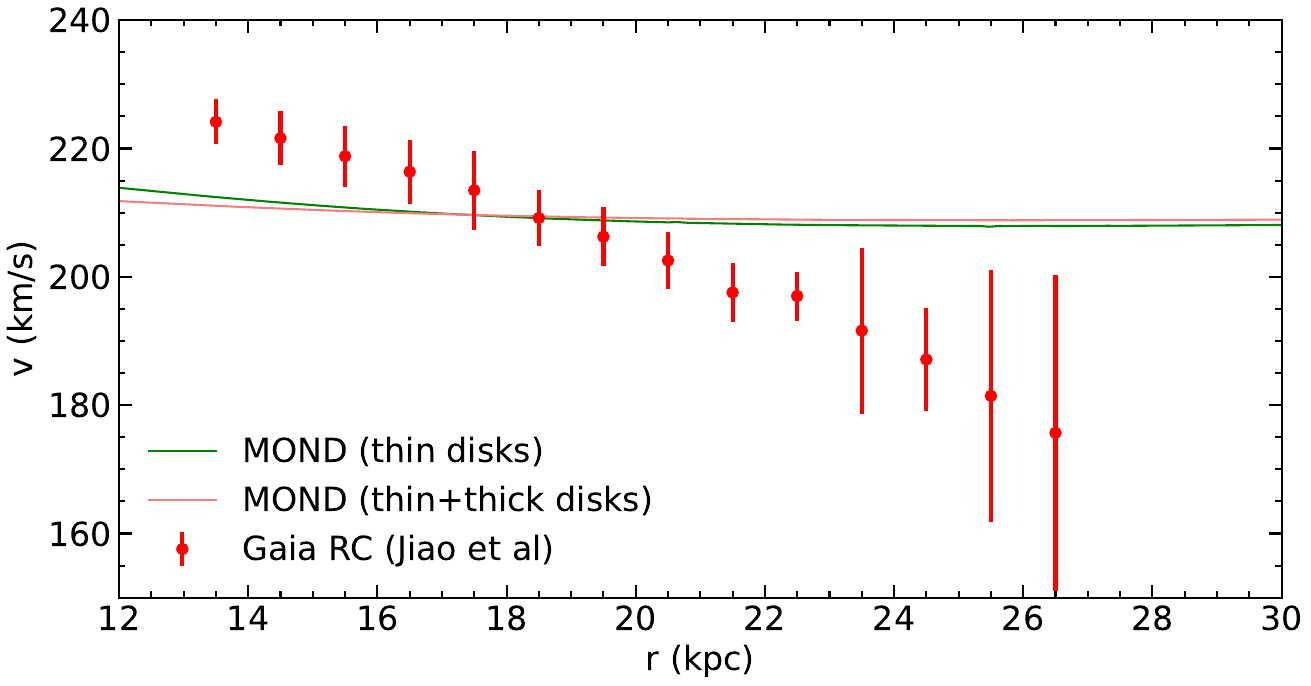}
    \caption{Decline fit under the MOND paradigm.  
    {{$a_0^{thin}=2.417\times10^{-10}$ m/s$^2$, $a_0^{thin+thick}=2.429\times10^{-10}$ m/s$^2$}}. The parameters for the baryonic components are the same than in Figure \ref{fig:nfw_decline}.}
    \label{fig:mond_classic}
\end{figure}

We then fit the declining part of the rotation curve  with this MOND paradigm, under the baryonic model{{s}} described above. The value of $a_0$ was determined by minimizing the $\chi^2$ considering the \cite{jiao_detection_2023} declining part of the RC. The best value of $a_0$ found here is $2.417\times10^{-10}$ m/s$^2$, which is not consistent with the value derived from the RC in other galaxies. The resulting RC is shown Figure \ref{fig:mond_classic}. {{The corresponding $\chi^2$, 49.7 and  57.6 (reduced $\chi^2$ of 3.83 and 4.43), confirm the visual impression that the fits are not satisfying.}}
{{We thus conclude that using the B2 model described as in \cite{jiao_detection_2023}, both with thin and thick disks,}} the declining rotation curve cannot be reproduced with  Milgrom dynamics.  {The physical reason of the behavior of the RC can be understood: the bulk of the mass is at lower distance than the  13 kpc. The mass is low so that above 13 kpc the dynamic is in the Mondian regime and the RC is nearly flat. }

\section{Relaxing the baryonic components to fit the decline with MOND}

\subsection{Methodology}

\noindent In the previous section, we have seen that the MOND theory could not accommodate the declining part of the rotation curve of the Milky Way, when the baryonic components are those of the B2 model in section \ref{s:bcmw}. In this section we examine whether relaxing the properties of the baryonic components could alleviate this inconsistency.
As the disk mass is the dominant baryonic component, it  is clearly an important parameter in featuring the Milky Way rotation curve. We thus treat this quantity as a free parameter, as well as $a_0$.  For the scale radius, we use two values : $r_d=2.35$ kpc (which is the value measured by \cite{misiriotis_distribution_2006} and used by \cite{jiao_detection_2023} as well as \cite{de_salas_estimation_2019}), and $r_d=3.1$ kpc, which are close to the bounds provided by \cite{bland-hawthorn_galaxy_2016}.

To determine our two parameters, the disk mass $M^{st.}$  and $a_0$, we ran two MCMC, one for each scale radius.  We apply the MCMC  Ensemble Slice Sampler algorithm from \textsc{zeus} \citep{2021MNRAS.508.3589K}. 
    $M^{st.}$ is allowed to vary between $3\times10^{10}$ M$_{\odot}$ and $2\times10^{11}$ M$_{\odot}$ and  $a_0$ between $0$ and $3\times10^{-10}$ m/s$^2$.

\subsection{Results}

\noindent Applying the methodology described above yields constraints on the parameter space, visualized by the contours in  Figure \ref{fig:mcmc_results}. The red cross indicates standard values of  $M^{st.}$ and $a_0$ and their uncertainties. The uncertainty on $M^{st.}$ is representative of the extreme values found in \cite{binney_galactic_2011} and \cite{bland-hawthorn_galaxy_2016}, whereas the uncertainty on $a_0$ can be found in \cite{milgrom_mond_2015}. One can notice that there is a correlation between the stellar disk mass and $a_0$. Moreover, the value of $r_d$ does not have much impact on the results. From these contours we extract a pair of values $(M^{st.},a_0)$ presented in Table \ref{tab:mcmc_results} that minimizes the $\chi^2$ :  {$\chi^2 = 5.53$ for $r_d=2.35$ kpc, and $\chi^2 = 3.84$ for $r_d=3.1$ kpc, slightly higher than with NFW, but which are still acceptable considering we only have two free parameters.

\begin{table}[ht]
    \centering
    \caption{Central values for $r_d = 2.35$ kpc and $r_d = 3.1$ kpc}
    \label{tab:mcmc_results}
    \begin{tabular}{ccc}
        \hline
		$r_d$ (kpc) & $M^{st.}$ (M$_{\odot}$) & $a_0$ (m/s$^2$) \\ 
		\hline
		$2.35$ & $\left( 11.01^{+0.63}_{-0.76} \right) \times 10^{10}$ & $\left( 0.65^{+0.12}_{-0.10} \right) \times 10^{-10}$ \\ 
		$3.1$ & $\left( 10.95^{+0.62}_{-0.66} \right) \times 10^{10}$ & $\left( 0.592^{+0.104}_{-0.092} \right) \times 10^{-10}$ \\ 
		\hline
    \end{tabular}
\end{table}

\begin{figure}[ht]
    \centering
    \includegraphics[width=0.9\linewidth]{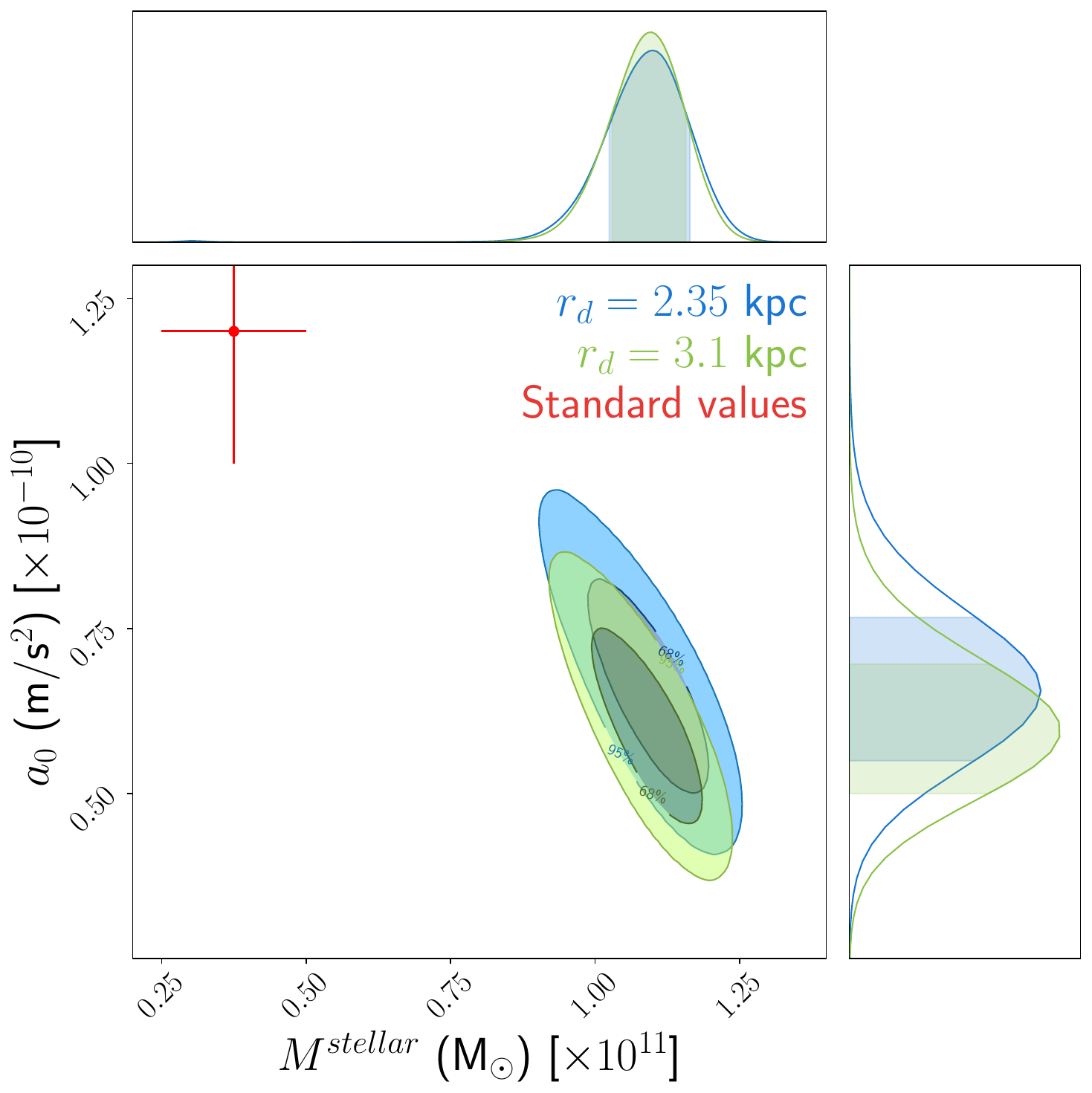}
    \caption{MCMC contours for $r_d=2.35$ kpc and $r_d=3.1$ kpc. The stellar disk mass and $a_0$ are left as free parameters. The red cross indicates the observed values of $M^{stellar}$ and the  {standard value of} $a_0$ with their respective uncertainty (see text).}
    \label{fig:mcmc_results}
\end{figure}

\begin{figure}[ht]
    \centering
    \includegraphics[width=1\linewidth]{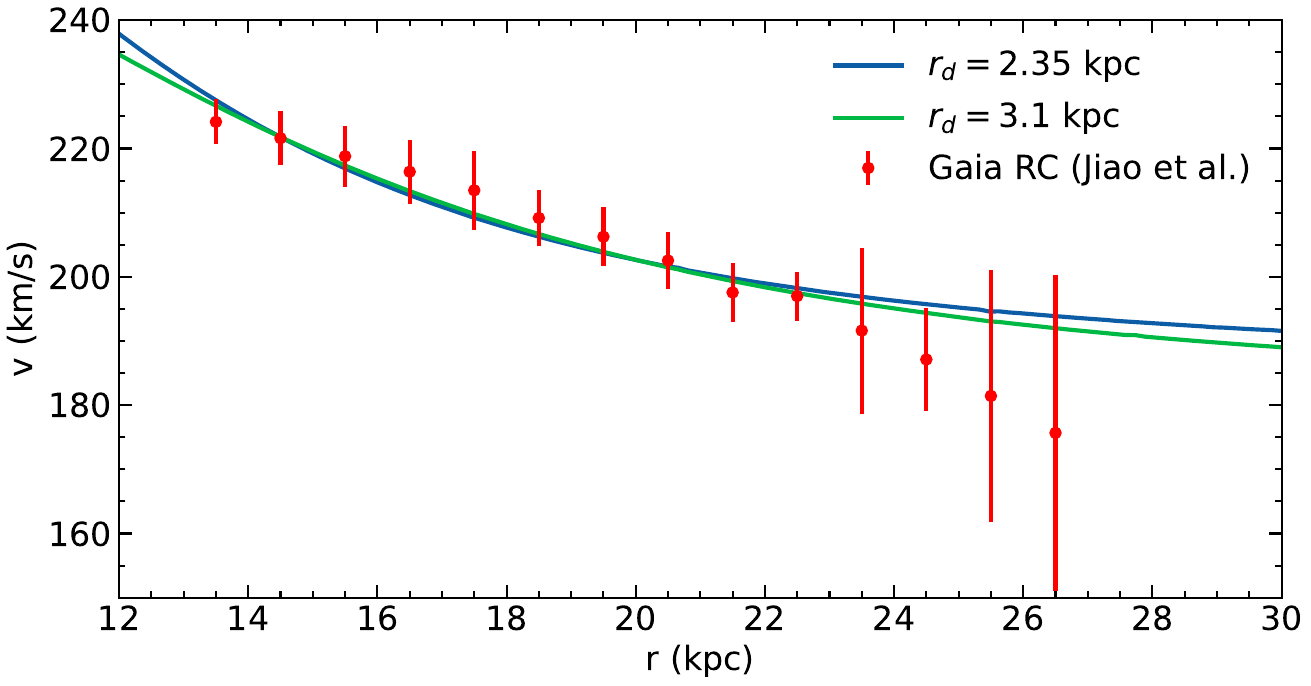}
    \caption{Fits using MCMC results  {(Table \ref{tab:mcmc_results}. The parameters for the other  baryonic components}  are the same as in Figure \ref{fig:mond_classic}.}
    \label{fig:mond_best_fits}
\end{figure}

As the red cross indicates, the obtained value of $a_0$ is smaller than the standard $1.2\times10^{-10}$ m/s$^2$ and is not consistent with previous results, namely from \cite{begeman_extended_1991}. In other words, according to our study, there is no way for MOND to explain both the Milky Way rotation curve and other galaxies' rotation curves with the same value of $a_0$.

Moreover, the stellar disk mass used to find such a value is about $10\times10^{10}$ solar masses, which is not consistent with the value $(3.5\pm1)\times10^{10}$ M$_{\odot}$ from \cite{bland-hawthorn_galaxy_2016}.  {The physical reason is that in order to produce a decreasing RC at $r>$ 15 kpc  the dynamic needs to be close to the Newtonian regime, needing a high mass (more than 95‰ of the mass lies within $r<$ 15 kpc). The Mondian regime starts at larger radius with a value of $a_0$  lower than the standard value. }

\subsection{More freedom on the baryonic components. }

\begin{figure*}[ht]
    \centering
    \includegraphics[width=0.8\linewidth]{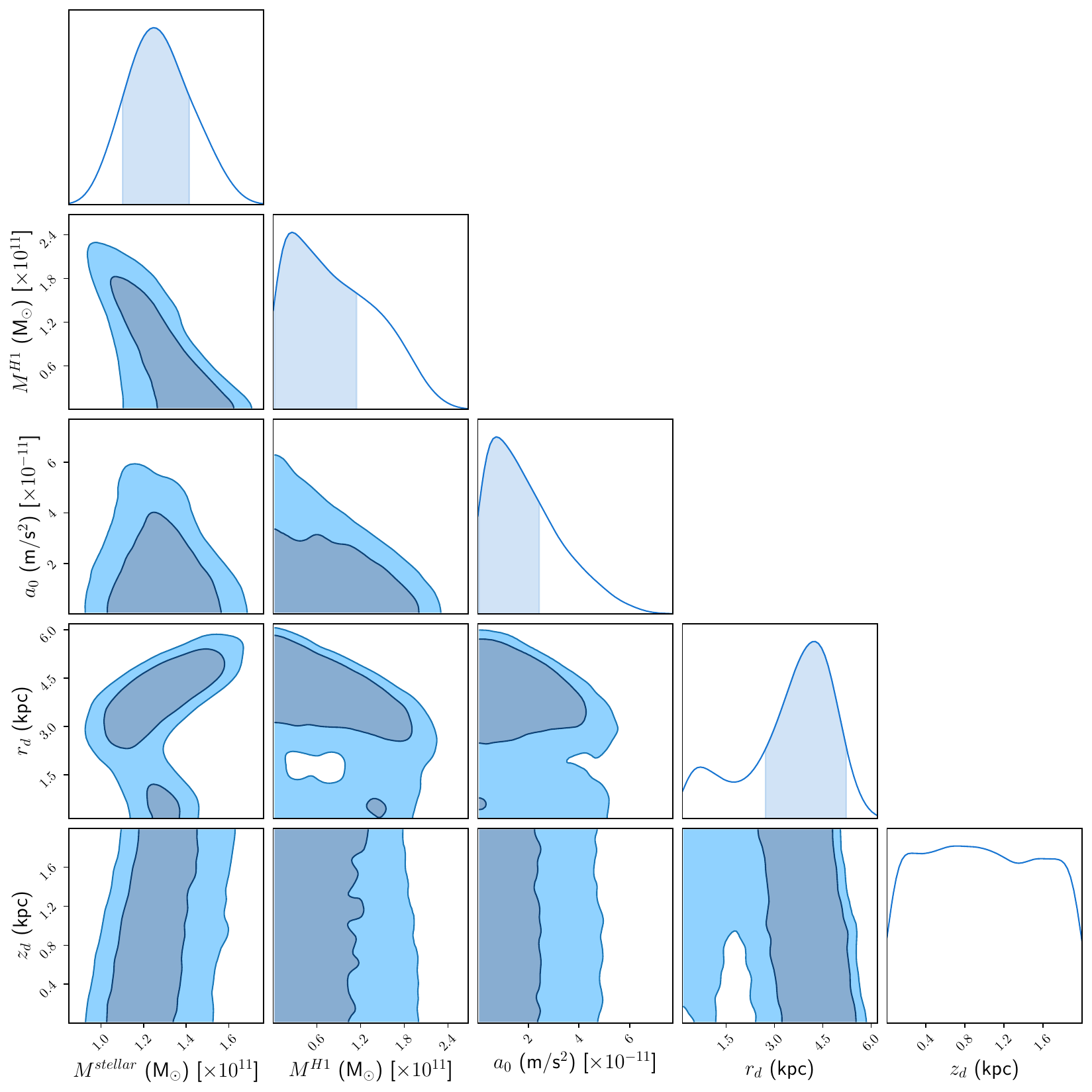}
    \caption{MCMC contours with a free stellar disk mass, free \text{H\RomanNumeralCaps{1}} disk mass, free $a_0$, free scale radius and free scale height.}
    \label{fig:mcmc_total}
\end{figure*}

\noindent Since relieving constraints on the stellar disk mass alone does not yield satisfying results, in this section we explore possibilities by increasing the number of free parameters on the baryonic components. We launch another MCMC with $M^{st.}$, $a_0$, $M^{\text{H\RomanNumeralCaps{1}}}$,  $r_d$ and $z_d$  as free parameters. Each prior used for this MCMC can be found in Table \ref{prior_mcmc}.

Figure \ref{fig:mcmc_total} reveals a number of interesting properties that can be inferred : it is possible to fit adequately the rotation curve with the baryonic components 
provided that they are allowed to take values well above the standard values~: for the models taken in the 1$\sigma$ domain, the $\chi^2$ typically lies between 3 and 6 which is acceptable and similar to the $\Lambda$CDM case. The scale height of the disk has no correlation with other parameters (except a weak correlation with the disk mass) and has no preferred value. The 
\text{H\RomanNumeralCaps{1}} disk mass $M^{\text{H\RomanNumeralCaps{1}}}$ has little to no impact on the results, as a wide range of values allows to fit the decline. Large values of the \text{H\RomanNumeralCaps{1}} disk mass are allowed, much above  the standard value, but remaining below the preferred stellar disk mass. The 1$\sigma$ intervals 
for the scale radius of the stellar disk and for its mass are higher than found in \cite{bland-hawthorn_galaxy_2016} and show a correlation with the stellar disk mass. Relieving the constraints on the baryonic matter distribution yields a lower value for $a_0$ than found in the previous MCMC analysis, and $a_0$ does not appear to be correlated with the baryonic distribution parameters. One can notice that $a_0=0$ m/s$^2$ is in the $1\sigma$ interval, assuming a heavy stellar disk and a scale radius $r_d>3$ kpc. The fact that the stellar disk is heavier than usual observations \citep{bland-hawthorn_galaxy_2016} can be explained by the close-to-zero value of $a_0$~: since little to no MONDian effect is requested and no dark halo is assumed, one needs extra matter in the disk to reach an acceptable value of the circular velocity. Essentially, this consists in a dark matter disk instead of the halo used in Figure \ref{fig:nfw_decline}, while models using a lighter disk and a non vanishing $a_0$ are not significantly preferred over the $a_0=0$ paradigm.

\begin{table}[ht]
    \centering
    \caption{Prior used for the MCMC Figure \ref{fig:mcmc_total}.}
    \begin{tabular}{ccc}
    \hline
    Parameter                   & Lower bound    & Higher bound    \\\hline 
    
    $M^{st.}$ (M$_{\odot}$) & $3\times 10^{10}$  & $2\times10^{11}$  \\  
    $a_0$ (m/s$^2$)             & 0              & $3\times10^{-8}$  \\
    $M^{gas}$ (M$_{\odot}$)     & 0              & $3\times10^{11}$   \\
    $r_d$ (kpc)                 & 0              & 20              \\
    $z_d$ (kpc)                 & 0           & 2   \\
    \hline 
    \end{tabular}
    \label{prior_mcmc}
\end{table}

\newpage
\section{Discussion and conclusion}

\noindent 
In this work, we examine the two major solutions to the missing mass problem in galaxies, applied to the Milky Way. More precisely, we compare the ability of $\Lambda$CDM (dark matter) and MOND (modified dynamics) to fit the Milky Way declining rotation curve measured by Gaia,  {assuming a simple dynamic that is not significantly disturbed in the outer parts of the Galaxy \citep{2024A&A...692A..50K,2024ApJ...970...94K}}.
Using a standard model for the baryonic components of the Milky Way we show that a simple dark matter distribution model like NFW, expected in $\Lambda$CDM, is able to explain Gaia's decline with ease, although with extreme value of the concentration parameter, whereas the MOND formulation cannot accommodate the decline under the B2 model, even when allowing $a_0$ to be a free parameter.
We then consider relieving constraints on baryonic parameters as well as the value of $a_0$ in order to examine whether MOND could accommodate the decline. For this we perform an MCMC on $a_0$ and the parameters of the baryonic components on the data of the declining part of the Milky Way rotation curve. Preferred models have disk masses at odds with values inferred from observations, with no significant preference for a non vanishing $a_0$.  We get an upper limit on  $a_0$ of $0.53\times10^{-10}$ m/s$^{-2}$ (95\%), significantly lower than what has been found necessary to fit flat rotation curves in other galaxies with MOND, nearly 5 $\sigma $ away. 

We conclude that the declining rotation curve of the Milky Way as recently inferred from  Gaia's data   {can be interpreted due to the presence of an NFW-type dark matter halo while not easy to reproduce in } the MOND alternative.

\section*{Acknowledgement}
This work was supported by the Programme National Cosmology et Galaxies (PNCG) of CNRS/INSU with INP and IN2P3, co-funded by CEA and CNES. This work was supported by CNES. 

\bibliographystyle{aa}

\bibliography{Biblio.bib}
\end{document}